\documentclass[aip,jcp,reprint,amsmath,amssymb]{revtex4-1}

   \usepackage[pdftex]{graphicx}
   \DeclareGraphicsExtensions{.pdf, .png, .jpg}
   \graphicspath{{./FIG/}}
   \usepackage{thumbpdf}
\usepackage{bm}
\usepackage{dcolumn}
\usepackage[yyyymmdd,hhmmss]{datetime}
\usepackage[colorlinks=true,urlcolor=blue,citecolor=blue,linkcolor=blue,breaklinks=true]{hyperref}
\usepackage{orcidlink}
\usepackage{physics}

\definecolor{cream}{RGB}{222,217,201}
\usepackage{multirow}
\usepackage{threeparttable}
\begin{document}

\title{Computing Shear Viscosities from Molecular Dynamics Simulation: Comparing the OrthoBoXY Approach with the Green-Kubo Method}

\author{Marcel Brandt\orcidlink{0009-0004-8691-4176}}
\email{marcel.brandt@uni-rostock.de}
\affiliation{Institut f\"ur Chemie, Abteilung Physikalische und Theoretische Chemie, 
Universit\"at Rostock, Albert-Einstein-Str.~27, D-18059 Rostock, Germany}

\author{Ralf Ludwig\orcidlink{0000-0002-8549-071X}}
%\email{ralf.ludwig@uni-rostock.de}
\affiliation{Institut f\"ur Chemie, Abteilung Physikalische und Theoretische Chemie, 
Universit\"at Rostock, Albert-Einstein-Str.~27, D-18059 Rostock, Germany}
\affiliation{Department Life, Light \& Matter, Universit\"at Rostock, 
Albert-Einstein-Str.~25, D-18059 Rostock, Germany}
\affiliation{Leibniz-Institut f\"ur Katalyse an der Universit\"at Rostock e.V., Albert-Einstein-Str.~29a, D-18059 Rostock, Germany}

\author{Dietmar Paschek\orcidlink{0000-0002-0342-324X}}
\email{dietmar.paschek@uni-rostock.de}
\affiliation{Institut f\"ur Chemie, Abteilung Physikalische und Theoretische Chemie, 
Universit\"at Rostock, Albert-Einstein-Str.~27, D-18059 Rostock, Germany}

\date{\today~at~\currenttime}

\begin{abstract}
We calculated shear viscosities of 15 neat molecular liquids from equilibrium
molecular dynamics (MD) simulations using the OrthoBoXY approach
 and compare them to viscosities calculated via the Green-Kubo method. 
While computing viscosities spanning three orders of magnitude,
we find that data obtained from both methods agree very well. 
%while discussing 
Here, we show how to avoid numerical pitfalls while computing the OrthoBoXY-data to obtain 
optimal results.
From simulations of multiple system sizes, we could verify that the viscosity of molecular liquids is not influenced by finite size effects down to systems as small as 250 molecules. 
%We have derived an equation for the standard error of the viscosity using the propagation of uncertainty.
Moreover, we demonstrate that also 
the computed standard error of the viscosity is 
nearly independent of the system size. This is shown to be a consequence
of a compensation effect of an increasing accuracy of the computed 
Einstein self-diffusion coefficients
with increasing systems-size and the system-size dependent weighting according to the OrthoBoXY-equation.
%We have derived an equation for the standard error of the viscosity using the propagation of uncertainty.
As a consequence, we suggest that it is preferable to run simulations of smaller systems with longer simulation times rather than larger systems with shorter simulation runs.
In addition, we discuss a refinement of the recently introduced ``recipe'' for OrthoBoXY simulations, which suggested a certain simulation block length $\tau_\mathrm{block}$ based on the average displacement of the particles. Based on data from simulations with varying run-lengths, we suggest the following modification: for highly viscous systems (with block lengths $\tau_\mathrm{block} > 100\,\textrm{ns}$), the value of $\tau_\mathrm{block}$ might safely be scaled by a factor of $1/8$, significantly reducing the computational resources needed. For less viscous systems ($1\,\textrm{ns} < \tau_\mathrm{block} < 100\,\textrm{ns}$), the value of $\tau_\mathrm{block}$ might safely be scaled by a factor of $1/4$, but caution is advised when going lower than that. 
For systems with high fluidity ($\tau_\mathrm{block} < 1\,\textrm{ns}$), the value of $\tau_\mathrm{block}$ should not be scaled down in order to achieve reliable results. 
When using a smaller system size of 250 molecules,
these refinements are 
leading up to a 24-fold reduction in computational cost
compared to the previous recommended set-up
without sacrificing numerical
accuracy.
%short simulations do not require significant computational resources, there is no need to sacrifice the accuracy of the data.
\end{abstract}

\maketitle

\section{Introduction}

The viscosity is a fundamental kinematic property of fluid systems with importance to many industrial applications. As a transport phenomenon, the shear viscosity describes the transport of momentum in a fluid. \cite{viscosity_of_liquids} It is defined by Newton's law of viscosity
\begin{equation}
\tau_{xy} = \eta \frac{\partial v_x}{\partial y},
\end{equation}
where $\tau$ is the shear stress, $\eta$ is the shear viscosity, and $\partial v_x / \partial y$ is the velocity gradient perpendicular to the flow direction. \cite{viscosity_of_liquids} Next to the shear viscosity, there is also the bulk viscosity which is however often of lesser interest. While the shear viscosity is related to the shear deformation of a system, the bulk viscosity is related to the compression or expansion of a system. \cite{landau_lifshitz} In this work, we will focus on
the shear viscosity. Hence, when referring to the viscosity, it is always implied to be the shear viscosity.

Next to experimental measurements, molecular dynamics (MD) simulations have become an important tool for determining the viscosity of a system. The ``classical'' way of calculating the viscosity from equilibrium MD simulations is the Green-Kubo method.\cite{green_1954,kubo_1957} It works by integrating 
the stress tensor autocorrelation function
\begin{equation}
\eta = \frac{V}{k_\textrm{B} T} \int_0^\infty \left< P_{ij}(0) \cdot P_{ij}(t) \right> \dd t\,,
\end{equation}
where $V$ is the volume, $k_\textrm{B}$ is Boltzmann's constant, $T$ is the temperature, and $P_{ij}$ is an off-diagonal ($i \neq j$) element of the stress tensor.\cite{maginnBestPracticesComputing2019} Another way of calculating the viscosity from equilibrium MD simulations is the recently
suggested OrthoBoXY approach.\cite{buschOrthoBoXYSimpleWay2023,busch_pccp2024,buschComputingAccurateTrue2024} This approach exploits the system-size dependence of self-diffusion coefficients 
in non-cubic systems to obtain the viscosity. A concise explanation of the OrthoBoXY approach is given in the second section of this paper.

Recently, a study was published by Smith and Sega \cite{smithInsightsVirtualChemistry2025,smith_erratum_2026} in which both the shear and bulk viscosities were calculated via the Green-Kubo method for a set of 146 neat molecular liquids using two popular force fields: GAFF and OPLS. In order to compare the shear viscosities from the OrthoBoXY approach and the Green-Kubo method, we have selected 15 ``common'' organic substances from this set, including alcohols, carbonyl compounds, ethers, aromatic compounds and nitriles. The liquids were chosen such that they cover a relatively wide range of viscosities (approximately $10^{-1}$ to $10^2$\,mPa\,s, or three orders of magnitude). Since the focus of this work is on the methods and not on the 
particular force fields, the selected systems were 
exclusively described via the OPLS forcefield.

In addition to just comparing numerical viscosity data, we also 
utilize selected systems to study how to 
further reduce the required computational
cost for determining viscosities via the OrthoBoXY approach. By studying
both, the variation of system size and simulation length, we can provide
improved simulation set-up recommendations for OrthoBoXY simulations, leading
up to a 24-fold reduction in computational cost without sacrificing numerical accuracy.

\section{Theoretical Background}

Self-diffusion coefficients $D$ calculated from MD simulations 
%with periodic boundary conditions
are well known to exhibit a system-size dependence. \cite{dunwegMolecularDynamicsSimulation1993} This is due to
the effect of periodic boundary conditions (PBCs) by inducing hydrodynamic
self-interactions in liquid systems. Derived from a hydrodynamic model, the relationship between the ``true'' self-diffusion coefficient $D_0$ and the self-diffusion coefficient from simulations with PBCs $D_\textrm{PBC}$ is described by the Yeh-Hummer equation \cite{yehSystemSizeDependenceDiffusion2004}
\begin{equation}
\label{yeh-hummer}
D_0 = D_\textrm{PBC} + \frac{k_\textrm{B} T \zeta}{6 \pi \eta L},
\end{equation}
where $k_\textrm{B}$ is Boltzmann's constant, $T$ is the temperature, $\zeta$ is a Madelung constant analogue, $\eta$ is the viscosity, and $L$ is the length of the simulation box. 
For cubic simulation boxes,
the Madelung constant analogue $\zeta$ 
%is a geometry-dependent property and
has a value of 
$2.8372\ldots$, which
%value of $\zeta$ 
can be calculated 
numerically via Ewald summation.\cite{yehSystemSizeDependenceDiffusion2004} 
The effect of PBCs on other transport properties has been reviewed recently.\cite{jamal_2020,celebi_2021,hulikalchakrapaniImpactFinitesizeEffects2025}
For non-cubic systems, the effect of PBCs on the diffusion has been shown to 
become direction-dependent.\cite{botan_2015,kikugwa_2015a,kikugawaHydrodynamicConsiderationFinite2015,voegele_2016,asta_2017}
For orthorhombic unit-cells, the direction-dependent self-diffusion coefficients  
can be expressed similar to the Yeh-Hummer formula\cite{buschOrthoBoXYSimpleWay2023}
\begin{equation}
D_0 = D_{\textrm{PBC},ii} + \frac{k_\textrm{B} T \zeta_{ii}}{6 \pi \eta L_i}\;,
\end{equation}
where $i$ denotes the $x$-, $y$- or $z$-direction. 
%The direction-dependent self-diffusion coefficient 
%$D_{\textrm{PBC},ii}$ will be referred to as $D_i$ for short. 
Here the exact values for $\zeta_{ii}$ depend on the geometry of the system \cite{buschOrthoBoXYSimpleWay2023,busch_pccp2024} and
can be determined via Ewald summation.\cite{orthoboxy} 
The OrthoBoXY approach \cite{buschOrthoBoXYSimpleWay2023} utilizes a specific geometry with the box length ratios $L_z / L_x = L_z / L_y = 2.7933\ldots$, where 
$\zeta_{xx}=\zeta_{yy}=0$ for the $x$- and $y$-direction. This means that the correction term vanishes and the simulated self-diffusion coefficient 
for those directions
are equal to the ``true'' self-diffusion coefficient,
leading to $D_0=(D_{\textrm{PBC},{xx}}+D_{\textrm{PBC},{yy}})/2$. 
In the $z$-direction however, the self-diffusion coefficient still shows a system-size dependence. This can be exploited 
to calculate the viscosity by rearranging the Yeh-Hummer-like equation with respect to the viscosity
\begin{equation}
\label{orthoboxy viscosity}
\eta = \frac{k_\textrm{B} T \zeta_{zz}}{6 \pi L_z (D_0 - D_{\mathrm{PBC},zz})}\;.
\end{equation}
Here $\zeta_{zz}$ has a value of 8.1711... 
when using the OrthoBoXY geometry.\cite{buschComputingAccurateTrue2024,orthoboxy} The idea of obtaining the viscosity from the system-size dependence 
of the self-diffusion coefficient had been originally proposed by Jamali \textit{et al.}. They have used multiple simulations of differently sized cubic boxes 
to calculate the viscosity.\cite{jamaliShearViscosityComputed2018} The advantage of the OrthoBoXY 
approach is that it is possible
%lies in the use of orthorhombic boxes with defined
%box-length ratios
to obtain both 
%the system-size independent diffusion coefficient 
$D_0$
and the viscosity $\eta$
from a single MD simulation run.

\section{Methods}

\begin{table}[t]
\caption{Overview of the investigated systems. Given are the temperature $T$, the density $\rho$, the thermostat used, and the simulation block length $\tau_\mathrm{block}$. A total of 40 subsequent simulation blocks were generated for each system. 
Reference densities are taken from Smith and Sega \cite{smithInsightsVirtualChemistry2025,smith_erratum_2026}.}
\centering
\label{systems}
\setlength{\tabcolsep}{0.105cm}
\begin{threeparttable}
\begin{tabular}{lcccc}
\hline\hline\\[-0.6em]
System & $T$/K & $\rho$/kg\,m$^{-3}$ & Thermostat & $\tau_\mathrm{block}$/ns\\[0.2em]
\hline\\[-0.6em]
Acetamide       & 300.00 & 1067.21          & V-rescale   & 18\tnote{b}\\
Acetone         & 298.15 &  801.02          & Nosé-Hoover & 0.7\\
Acetonitrile    & 293.15 &  761.54          & Nosé-Hoover & 0.5\\
Dimethyl ether  & 240.00 &  741.62          & Nosé-Hoover & 0.55\\
Ethanol         & 298.15 &  796.36          & Nosé-Hoover & 1.9\\
Formaldehyde    & 300.00 &  703.58          & Nosé-Hoover & 0.2\tnote{b}\\
Formamide       & 298.15 & 1218.85\tnote{a} & Nosé-Hoover & 7\\
Formic acid     & 293.15 & 1141.96          & Nosé-Hoover & 1.1\\
Glycerol        & 320.00 & 1235.55          & V-rescale   & 320\tnote{b}\\
3-Methylphenol  & 300.00 & 1042.84          & V-rescale   & 82.5\\
1-Pentanol      & 298.15 &  811.88          & V-rescale   & 9.6\\
Phenol          & 318.15 & 1060.81          & V-rescale   & 27\\
Pyridine        & 293.15 &  984.73          & Nosé-Hoover & 2.5\tnote{b}\\
Tetrahydrofuran & 298.15 &  858.86          & Nosé-Hoover & 1.3\\
Toluene         & 298.15 &  874.93          & Nosé-Hoover & 2.55\\[0.6em]
\hline\hline
\end{tabular}
\begin{tablenotes}
\footnotesize
\item[a]{For formamide, no OPLS density was given. The GAFF density was used instead.}
\item[b]{For acetamide, formaldehyde, glycerol and pyridine, additional simulations with lengths of $\frac{1}{2} \tau_\mathrm{block}$, $\frac{1}{4} \tau_\mathrm{block}$ and $\frac{1}{8} \tau_\mathrm{block}$ were performed. For glycerol, the $1 \tau_\mathrm{block}$ simulation was not performed.}
\end{tablenotes}
\end{threeparttable}
\end{table}

The investigated systems were modelled using the OPLS-AA force field \cite{jorgensenDevelopmentTestingOPLS1996}. The molecular geometry files and the topology files were obtained from a benchmark study by Caleman \textit{et al.} \cite{calemanForceFieldBenchmark2012}, available on \texttt{https://virtualchemistry.org}. The initial configurations with 1000 respective molecules in the OrthoBoXY geometry ($L_z / L_x = L_z / L_y = 2.7933...$) were created with PACKMOL \cite{martinezPACKMOLPackageBuilding2009}. For acetonitrile, ethanol and 1-pentanol, additional configurations with 250, 500 and 750 molecules were created. The simulations were performed 
in the NVT ensemble under PBCs
with an integration timestep of 2\,fs
using GROMACS 2019.6.\cite{abrahamGROMACSHighPerformance2015, lindahlGROMACS20196Source2020} 
Each system has an individual temperature and a volume corresponding to the density at the respective temperature as reported by Smith and Sega \cite{smithInsightsVirtualChemistry2025}. In general, the Nosé-Hoover thermostat \cite{noseMolecularDynamicsMethod1984, hooverCanonicalDynamicsEquilibrium1985} was used with a coupling constant of 2\,ps. However, some systems have shown a strong energy oscillation when using this algorithm. In these cases, Bussi's velocity rescale (V-rescale) thermostat \cite{bussiCanonicalSamplingVelocity2007} with a coupling constant of 2\,ps was used instead. Both the long-range Coulomb and Lennard-Jones interactions were calculated with the particle mesh Ewald (PME) summation using a cut-off radius of 1.1\,nm and an Ewald convergence parameter of $10^{-5}$ and $10^{-3}$, respectively. The Fourier spacing was set to one tenth of the cut-off radius. First, the systems were equilibrated for 2\,ns. Afterwards, the production runs were performed with varying simulation lengths. The respective temperature, density, thermostat, and simulation block length for each system can be found in Table~\ref{systems}.

The coordinates of all atoms were extracted from the simulations with different frequencies for each system, such that 2500 frames per simulation block were extracted. The mean squared displacement 
%$\left< \Delta r^2(t) \right>$ 
was calculated using the Python packages MDAnalysis \cite{michaud-agrawalMDAnalysisToolkitAnalysis2011, gowersMDAnalysisPythonPackage2016} and MDorado. % (\texttt{https://github.com/Paschek-Lab/MDorado}). 
The self-diffusion coefficients were calculated from the mean squared displacement according to the Einstein method
\begin{equation}
D_\textrm{PBC}
=\frac{1}{6}
\lim_{t\rightarrow\infty}
\frac{\dd}{\dd t}
\left<
|\mathbf{r}(0)
-
\mathbf{r}(t)
|^2
\right>\;,
\end{equation}
and
\begin{equation}
\label{einstein method}
D_{\textrm{PBC},{ii}}
=\frac{1}{2}
\lim_{t\rightarrow\infty}
\frac{\dd}{\dd t}
\left<
|r_{i}(0)
-
r_{i}(t)
|^2
\right>\;,
\end{equation}
where $\mathbf{r}(t)=[r_x(t),r_y(t),r_z(t)]$ represent the position of the center of mass
of a molecule at time $t$ and the $r_{i}(t)$ are its respective components
in $x$-, $y$-, and $z$-direction.
%\begin{equation}
%\label{einstein method}
%D_\mathrm{PBC,ii} = \frac{1}{2} \lim_{t \rightarrow \infty} \frac{\textrm{d}}{\textrm{d} t} \left< \Delta r^2(t) \right>,
%\end{equation}
%where $n_\textrm{D}$ is the number of dimensions. 
\cite{maginnBestPracticesComputing2019} In practice, the time derivative of the mean squared displacement was calculated as the slope of a linear fit from $0.03\times\tau_\mathrm{block}$ to 
$0.4\times\tau_\mathrm{block}$. 
%Note that $n_\textrm{D} = 1$ when determining the self-diffusion coefficient for the $x$-, $y$- and $z$-dimension individually, therefore the slope is multiplied by a factor of $\frac{1}{2}$ instead of $\frac{1}{6}$. 
The viscosity was then calculated from the difference between the ``true'' self-diffusion coefficient 
$D_0=(D_{\textrm{PBC},{xx}}+D_{\textrm{PBC},{yy}})/2$ 
%(from the $x$- and $y$-direction) 
and the system-size dependent self-diffusion coefficient 
$D_{\textrm{PBC},{zz}}$
%(from the $z$-direction) 
according to Eq.~\ref{orthoboxy viscosity}.

\section{Results and Discussion}
\subsection{Comparison of OrthoBoXY and Green-Kubo Viscosities}

\begin{table*}[t]
\caption{Viscosities of the investigated systems from the OrthoBoXY approach (this work) and the Green-Kubo method (Smith and Sega \cite{smithInsightsVirtualChemistry2025,smith_erratum_2026}) as well as experimental data from literature (references behind the individual values). The uncertainties reported for the OrthoBoXY and Green-Kubo viscosities are the standard errors, corresponding to a $1 \sigma$ confidence interval.}
\label{viscosity table}
\centering
\setlength{\tabcolsep}{0.91cm}
\begin{tabular}{lcccc}
\hline\hline\\[-0.6em]
\multirow{2}{*}{System} & \multirow{2}{*}{$T$/K} & \multicolumn{3}{c}{$\eta$/mPa\,s}\\
\cline{3-5}
& & OrthoBoXY & Green-Kubo & Experimental \\[0.2em]
\hline\\[-0.6em]
Acetamide       & 300.00 &  10.0 $\pm$ 0.8   &   15 $\pm$ 1    & 14.18 \cite{yaws2003yaws}\\
Acetone         & 298.15 & 0.318 $\pm$ 0.023 & 0.33 $\pm$ 0.04 & 0.31 \cite{haynes2016crc}\\
Acetonitrile    & 293.15 & 0.309 $\pm$ 0.021 & 0.29 $\pm$ 0.01 & 0.37 \cite{katritzkyPredictionLiquidViscosity2000}\\
Dimethyl ether  & 240.00 & 0.234 $\pm$ 0.022 & 0.23 $\pm$ 0.01 & 0.216 \cite{wuViscositySaturatedLiquid2003}\\
Ethanol         & 298.15 &  0.96 $\pm$ 0.08  & 0.93 $\pm$ 0.03 & 1.07 \cite{haynes2016crc}\\
Formaldehyde    & 300.00 & 0.160 $\pm$ 0.016 & 0.14 $\pm$ 0.02 & 0.11 \cite{yaws2003yaws}\\
Formamide       & 298.15 &   5.9 $\pm$ 0.5   & ---             & 3.23 \cite{katritzkyPredictionLiquidViscosity2000}\\
Formic acid     & 293.15 &  0.83 $\pm$ 0.07  & 0.90 $\pm$ 0.06 & 1.88 \cite{yaws2003yaws}\\
Glycerol        & 320.00 &   230 $\pm$ 24    &  160 $\pm$ 2    & 202.2 \cite{ferreiraViscosityGlycerol2017}\\
3-Methylphenol  & 300.00 &    70 $\pm$ 8     &   40 $\pm$ 2    & 6.61 \cite{rosalViscositiesDensitiesBinary2003, yasminDensityViscosityVelocity2011}\\
1-Pentanol      & 298.15 &  3.45 $\pm$ 0.28  &  3.3 $\pm$ 0.3  & 3.556 \cite{yangDensityViscosityBinary2006}\\
Phenol          & 318.15 &   9.1 $\pm$ 0.8   &  7.7 $\pm$ 0.5  & 4.00 \cite{guindaViscosityMeasurementsAniline1986}\\
Pyridine        & 293.15 &  1.11 $\pm$ 0.11  &  1.2 $\pm$ 0.1  & 0.97 \cite{katritzkyPredictionLiquidViscosity2000}\\
Tetrahydrofuran & 298.15 &  0.63 $\pm$ 0.05  & 0.53 $\pm$ 0.01 & 0.46 \cite{haynes2016crc}\\
Toluene         & 298.15 &  0.92 $\pm$ 0.08  & 0.80 $\pm$ 0.01 & 0.56 \cite{haynes2016crc}\\[0.6em]
\hline\hline
\end{tabular}
\end{table*}

\begin{figure*}[t]
	\includegraphics[width=0.8\textwidth]{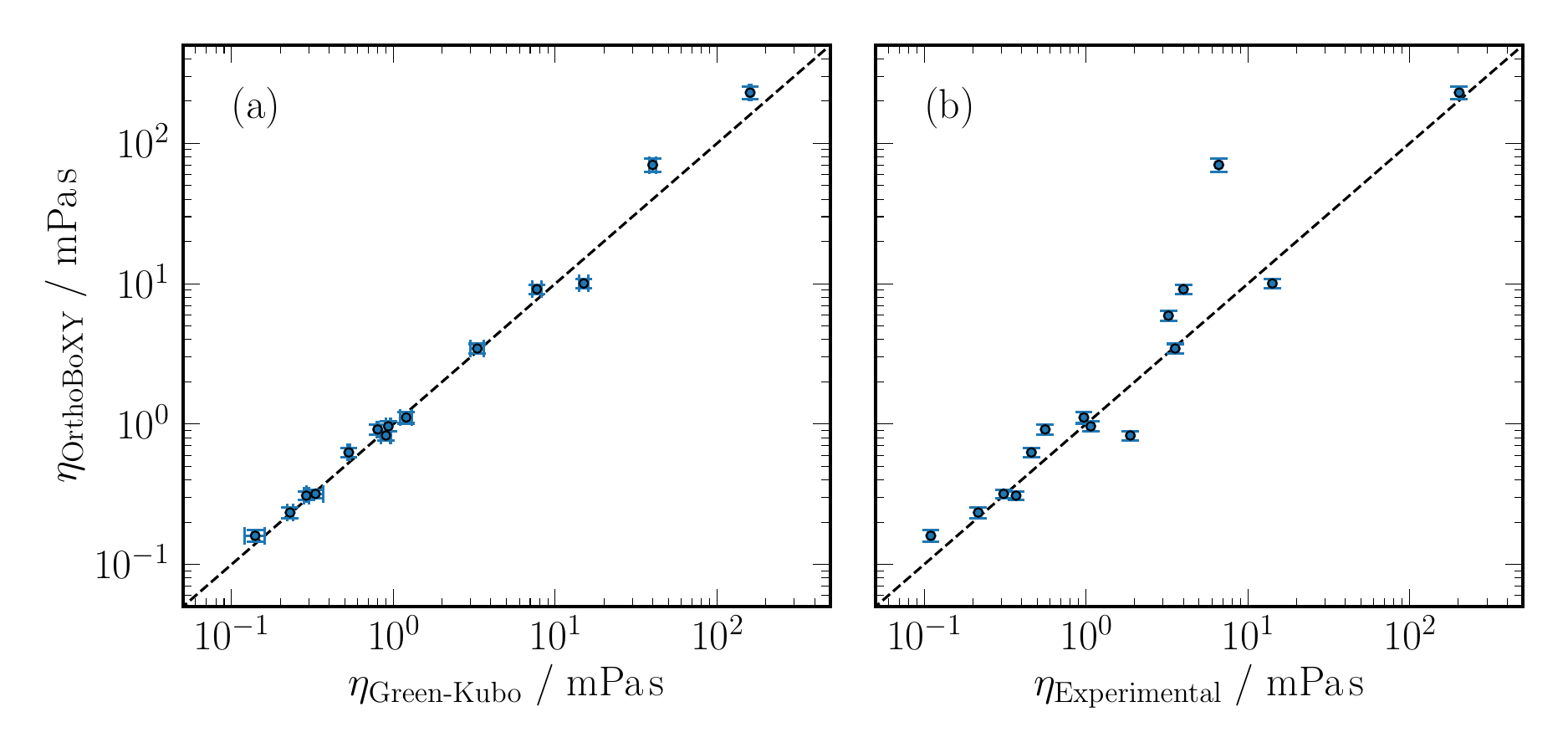}
	\caption{Comparison of OrthoBoXY viscosities with (a) Green-Kubo viscosities and (b) experimental viscosities from literature. For details on the data, see Table~\ref{viscosity table}.}
	\label{viscosity comparison}
\end{figure*}

The OrthoBoXY viscosities of the investigated systems are reported in 
Table~\ref{viscosity table}. All values are for 1000 molecules and 40 simulation blocks of length $\tau_\mathrm{block}$ (except for glycerol, where $\frac{1}{2} \tau_\mathrm{block}$ was used). Next to the OrthoBoXY viscosities, the Green-Kubo viscosities are 
provided (taken from Smith and Sega \cite{smithInsightsVirtualChemistry2025}). Additionally, experimental viscosities from literature are given (as compiled by Smith and Sega). Where no experimental viscosity for the specific temperature was found, a linear interpolation or empirical fit was used (see Ref.~\cite{smithInsightsVirtualChemistry2025}). The data is also graphically displayed in Fig.~\ref{viscosity comparison}. The left panel compares the OrthoBoXY viscosities with the Green-Kubo viscosities and the right panel with the experimental viscosities. As the Fig.~\ref{viscosity comparison}(a) shows, the OrthoBoXY and Green-Kubo viscosities agree quite well. The only noteworthy deviation can be seen for the three systems with the largest viscosities. For acetamide, the OrthoBoXY viscosity is smaller than the Green-Kubo viscosity, and for 3-methylphenol and glycerol, the OrthoBoXY viscosity is larger. From this data, it can not be said which method is responsible for the deviation, although it should be noted that the Green-Kubo method has been described by Maginn \textit{et al.} \cite{maginnBestPracticesComputing2019} as working best for relatively low viscosities ($<$ 20\,mPa\,s). The OrthoBoXY and experimental viscosities 
shown in Fig.~\ref{viscosity comparison}(b)
agree less well. This is however expected, as this comparison also involves the quality of the force field.

\subsection{Influence of the Statistical Treatment}

The MD simulations are split into multiple simulations blocks
of a certain length $\tau_\mathrm{block}$. Here the block lengths
are chosen such that the properties computed for subsequent blocks
are statistically independent from one another.
%. The properties of the system are often calculated for each simulation block separately. 
The individual results can then be averaged and the error can be estimated from the variance. Since the viscosity
is determined from direction-dependent self-diffusion 
coefficients, it is worthwhile to analyse various ways to
average the data to avoid numerical artifacts:

\textbf{Variant A:}
The perhaps obvious variant would be to calculate the difference of the self-diffusion coefficients $D_0 - D_{\mathrm{PBC},zz}$ for each simulation block separately and then average over the individual viscosity values
\begin{equation}
\eta = \left< \frac{k_\textrm{B} T \zeta_{zz}}{6 \pi L_z (D_0 - D_{\mathrm{PBC},zz})} \right>\;.
\end{equation}

\textbf{Variant B:}
For reasons discussed later, it might be beneficial to calculate the average of $D_0$ first. When assuming the fluctuations of $D_0$ and $D_{\mathrm{PBC},zz}$ are independent from each other, we can then calculate the difference $\left< D_0 \right> - D_{\mathrm{PBC},zz}$ for each simulation block
\begin{equation}
\eta = \left< \frac{k_\textrm{B} T \zeta_{zz}}{6 \pi L_z \left( \left< D_0 \right> - D_{\mathrm{PBC},zz} \right)} \right>\;.
\end{equation}

\textbf{Variant C:}
A third variant would be to calculate the average of both $D_0$ and $D_{\mathrm{PBC},zz}$ and then calculate the difference $\left< D_0 \right> - \left< D_{\mathrm{PBC},zz} \right>$. Obviously, no averaging of the individual viscosities is possible here, as only one value is produced
\begin{equation}
\eta = \frac{k_\textrm{B} T \zeta_{zz}}{6 \pi L_z \left( \left< D_0 \right> - \left< D_{\mathrm{PBC},zz} \right> \right)}\;.
\end{equation}

\begin{figure*}[t]
	\centering
    \includegraphics[width=0.8\textwidth]{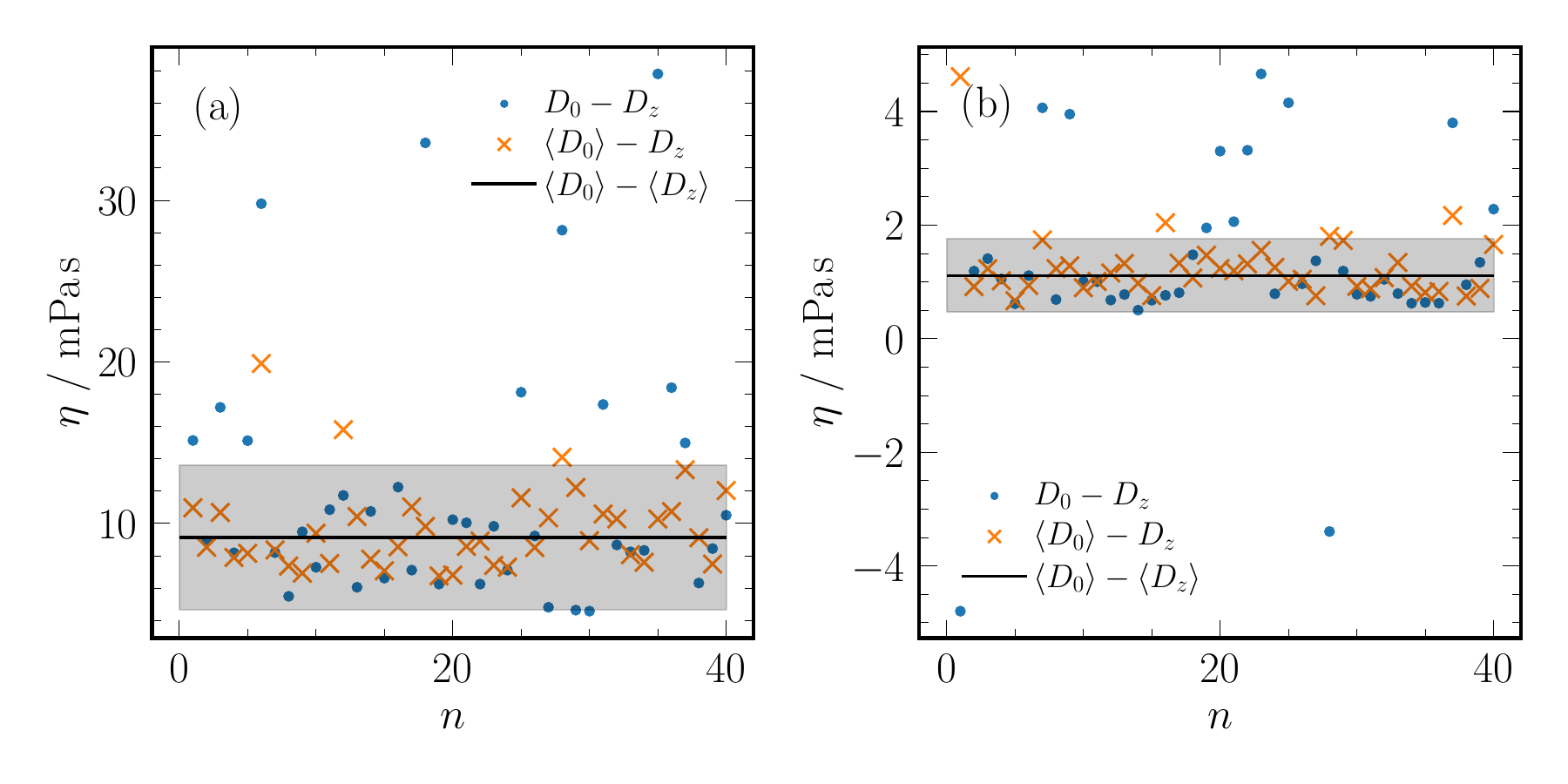}
	\caption{Viscosities of (a) phenol and (b) pyridine for the $n$th simulation block calculated via variant A (blue circles), variant B (orange crosses) and variant C (black line). Note that variant C gives only one value and is displayed with its standard deviation (shaded region).}
	\label{influence of stat}
\end{figure*}

In Fig.~\ref{influence of stat}, the viscosities of phenol and pyridine are calculated with these three variants. It can be seen, especially for variant A, that the viscosities do not follow a normal distribution. This is due to the viscosity being inversely proportional to the $D_0 - D_{\mathrm{PBC},zz}$ term, skewing the viscosity distribution to higher values. When the distributions of $D_0$ and $D_{\mathrm{PBC},zz}$ (which are assumed to be normal) have a significant overlap, it might even happen that $D_{\mathrm{PBC},zz}$ is higher than $D_0$, giving a negative viscosity value. In the limit of an infinite sample size, the negative values are just part of the distribution function. With a finite sample size however, these negative values can significantly skew the average and the error estimate. Therefore, we discourage the use of variant A.

Since the negative values result from an overlap of the distributions of $D_0$ and $D_{\mathrm{PBC},zz}$, it should be possible to avoid this problem by using the average of $D_0$ instead of the individual values (variant B). We use the average of $D_0$ (instead of $D_{\mathrm{PBC},zz}$) because it shows a larger variance. This approach noticeably reduces the scattering of the viscosity while still giving individual results for each simulation block. In most cases, variant B should give good results.

Variant C uses the average of both $\langle D_0\rangle$ and $\langle D_{\mathrm{PBC},zz}\rangle$. It is the most robust approach and should produce accurate results even when the difference between $\langle D_0\rangle$ and 
$\langle D_{\mathrm{PBC},zz}\rangle$ is very small. Since this approach produces only one value, the variance of the viscosity $\sigma_\eta^2$ can not be estimated from the individual values. We instead 
need to use the propagation of uncertainty to calculate $\sigma_\eta^2$ from the variance of the self-diffusion coefficients $\sigma_{D_0}^2$ and $\sigma_{D_{\mathrm{PBC},zz}}^2$.
\begin{equation}
\sigma_\eta^2 = \left( \frac{\partial \eta}{\partial D_0} \right)^2 \sigma_{D_0}^2 + \left( \frac{\partial \eta}{\partial D_{\mathrm{PBC},zz}} \right)^2 \sigma_{D_{\mathrm{PBC},zz}}^2
\end{equation}
From this equation and Eq.~\ref{orthoboxy viscosity}, we can find an expression for the absolute standard error $\hat{\sigma} = \sigma / \sqrt{N}$.
\begin{equation}
\label{absolute standard error}
\hat{\sigma}_\eta = \frac{k_\textrm{B} T \zeta_{zz}}{6 \pi L_z (D_0 - D_{\mathrm{PBC},zz})^2} \sqrt{\hat{\sigma}_{D_0}^2 + \hat{\sigma}_{D_{\mathrm{PBC},zz}}^2}
\end{equation}
Alternatively, an expression for the relative standard error can be found.
\begin{equation}
\label{relative standard error}
\frac{\hat{\sigma}_\eta}{\eta} = \frac{\sqrt{\hat{\sigma}_{D_0}^2 + \hat{\sigma}_{D_{\mathrm{PBC},zz}}^2}}{\left|D_0 - D_{\mathrm{PBC},zz}\right|}.
\end{equation}
All viscosities reported in this paper were calculated via variant C and the standard errors were estimated via error propagation.

\subsection{Influence of the System Size}

\begin{figure}[t]
	\centering
	\includegraphics[width=0.45\textwidth]{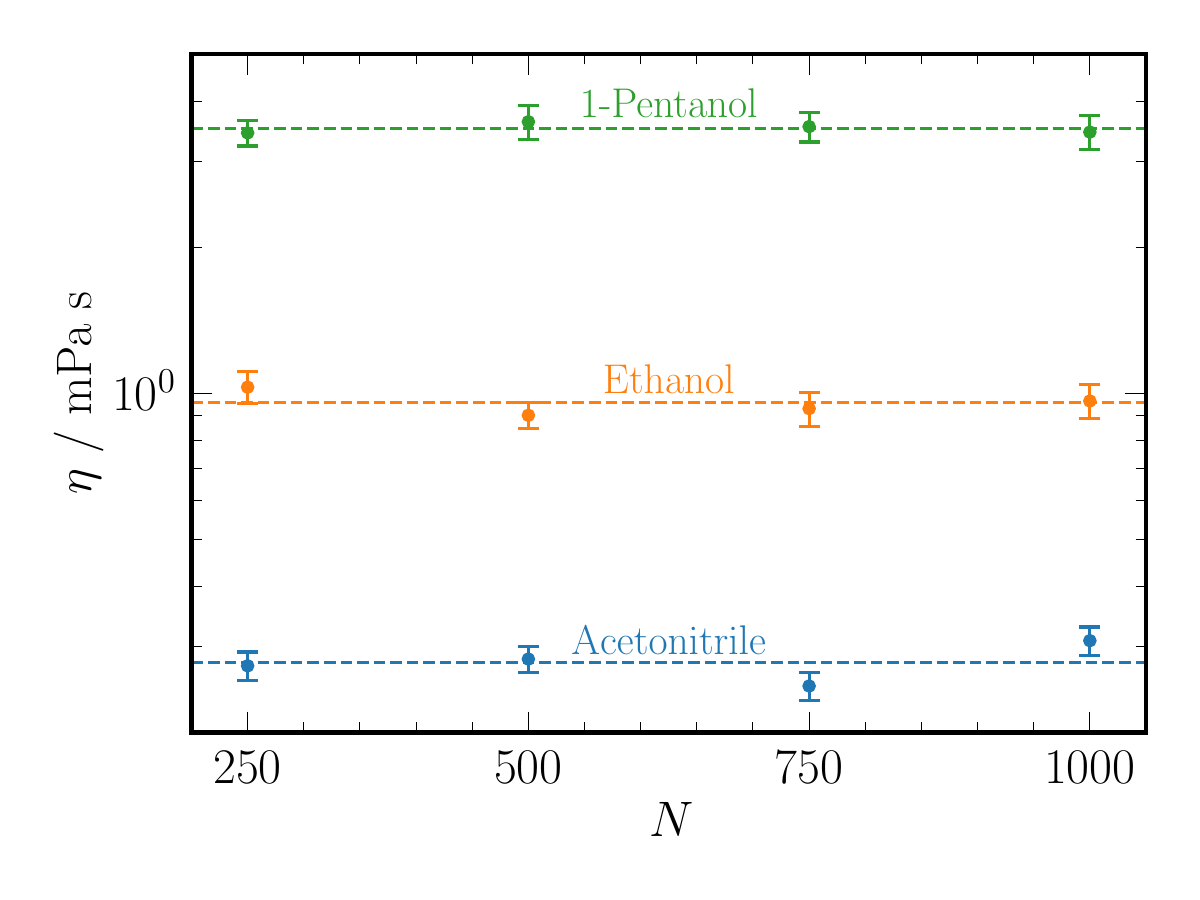}
	\caption{Viscosities of acetonitrile (blue), ethanol (orange) and 1-pentanol (green) for system sizes of 250, 500, 750 and 1000 molecules. The dashed lines represent the averages.}
	\label{influence of n}
\end{figure}
\begin{figure*}[t]
	\centering
	\includegraphics[width=0.95\textwidth]{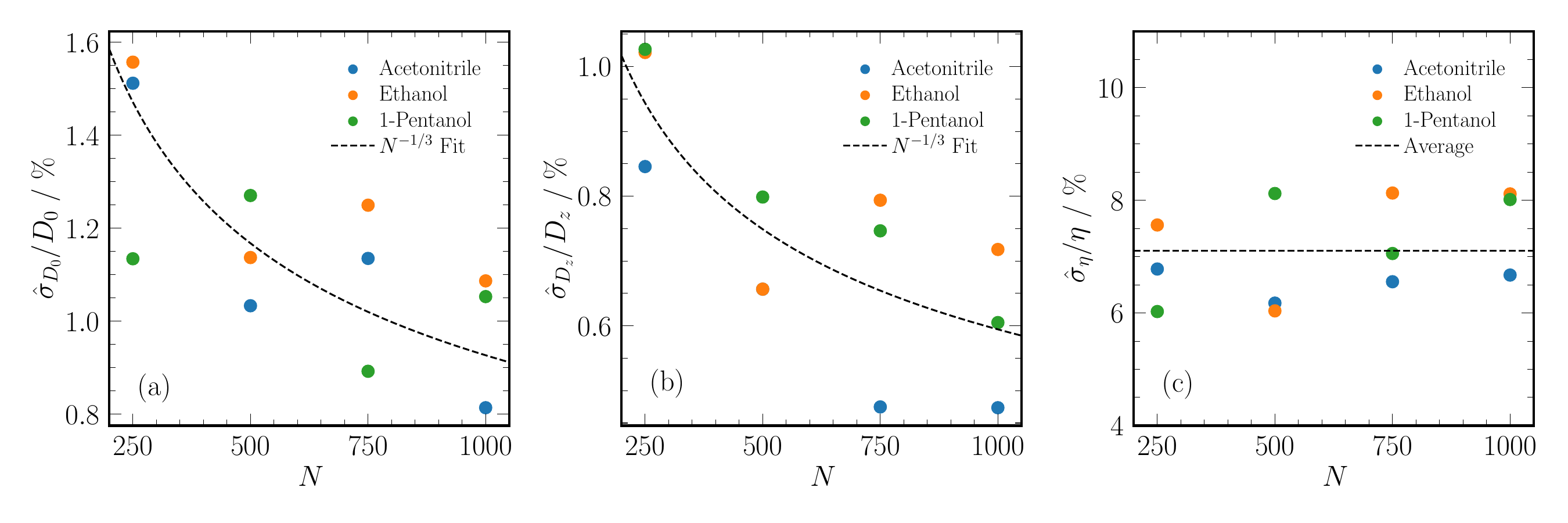}
	\caption{Relative standard errors of (a) $D_0$, (b) $D_{\mathrm{PBC},zz}$ and (c) $\eta$ computed for varying system sizes. The data are shown as a function of the number of molecules $N$ in the system.}
	\label{relative errors plot}
\end{figure*}

In contrast to self-diffusion coefficients, the viscosity of dense fluids 
generally does not 
depend on the system size.\cite{yehSystemSizeDependenceDiffusion2004, moultosSystemsizeCorrectionsSelfdiffusion2016} For very small systems, however, a system-size dependence is observed, where the viscosity experiences a pronounced oscillation as a function of system-size that 
gets dampened relatively quickly.\cite{kimNatureIntrinsicUncertainties2018, hulikalchakrapaniImpactFinitesizeEffects2025} Due to the dampening, this effect becomes negligible for sufficiently large system sizes. The amplitude of the oscillation and the point of convergence seem to be 
system-specific. Here simpler systems (e.g. Lennard-Jones fluids) are showing a stronger oscillatory behavior than complex systems. This effect is attributed to configurational rearrangements being not properly represented in small systems.

In addition, the system size may also affect the statistical accuracy of the computed viscosity data. In the framework of the Green-Kubo method, the standard deviation and consequently the precision of the computed 
viscosity is found to be largely independent of the system size.\cite{zhangReliableViscosityCalculation2015} 
This is due to the stress tensor being a property of the ensemble rather than a property of the individual particles. In the OrthoBoXY framework, the viscosity is ultimately calculated as an average over particle trajectories. Naively, 
one would therefore expect a system-size dependence.\cite{zwanzigStatisticalErrorDue1969}
%\begin{figure}[t]
%	\centering
%	\includegraphics[width=0.4\textwidth]{influence_of_n.pdf}
%	\caption{Viscosities of acetonitrile (blue), ethanol (orange) and 1-pentanol (green) with system sizes of 250, 500, 750 and 1000 molecules. The dashed lines represent the averages.}
%	\label{influence of n}
%\end{figure}
In our previous studies, however, essentially no significant system-size dependence of the computed
accuracy could be detected.\cite{buschOrthoBoXYSimpleWay2023,buschComputingAccurateTrue2024}
To elucidate this effect, we study the
system-size dependence of the
viscosities and their uncertainties 
of three selected systems: acetonitrile, ethanol, and 1-pentanol.

In Fig.~\ref{influence of n}, the viscosities of 
three selected systems are shown for four different system sizes (250, 500, 750, and 1000 molecules). No dependence  of the viscosity with the system size can be observed, as is expected for dense fluids. Also, an oscillatory behavior of the viscosity that goes beyond the statistical fluctuation cannot be observed. This is likely 
related to the complexity of the systems (compared to e.g. Lennard-Jones fluids). 
We would like to point out that the systems are modelled with an all-atom force field using flexible bonds, leading to many internal degrees of freedom which 
allow for configurational rearrangements.

%The assumed scaling of the error with $N^{-1/2}$ predicts the error for 1000 molecules to be half the error for 250 molecules. This relation can not be verified for any of the three systems. 
In accordance with our earlier observations\cite{buschOrthoBoXYSimpleWay2023,buschComputingAccurateTrue2024}, 
Figs.~\ref{influence of n} and \ref{relative errors plot}(c) indicate that
no clear correlation between the system size and the error 
of the computed viscosity
can be observed. 
From Eq.~\ref{relative standard error} follows
that the error of the viscosity depends on the error of the self-diffusion coefficients $D_0$ and $D_{\mathrm{PBC},zz}$ (see Fig.\ref{relative errors plot}(a) and \ref{relative errors plot}(b)), but it is weighted 
by the absolute
difference $|D_0-D_{\mathrm{PBC},zz}|$. 
In order to fulfill Eq.~\ref{orthoboxy viscosity}, however,
$|D_0-D_{\mathrm{PBC},zz}|$ has to be proportional to $L_z^{-1}$, and 
is therefore
scaling with the system-size via $N^{-1/3}$.
This weighting is counter-balancing the 
system-size dependent error of the computed self-diffusion coefficients.
Naively, one would expect that the error of the self-diffusion coefficient 
is scaling with $N^{-1/2}$. However, since the particle trajectories are not
completely indepdent from one another,
a scaling closer to $N^{-1/3}$ has been observed
for molecular liquids in earlier studies by Jamali \textit{et al.} \cite{jamaliShearViscosityComputed2018} and by us\cite{buschComputingAccurateTrue2024}. 
%This may indicate that the trajectories of the particles are not perfectly independent. The difference between $D_0$ and $D_{\mathrm{PBC},zz}$ is, according to the Yeh-Hummer equation (Eq.~\ref{yeh-hummer}), proportional to $L^{-1}$ or $N^{-1/3}$ as well. When applying these scalings to 
According to Eq.~\ref{relative standard error}, apparently
both effects exactly cancel out each other, giving a viscosity error independent of $N$. In order to test this prediction, we have displayed the relative standard errors of $D_0$, $D_{\mathrm{PBC},zz}$, and $\eta$ in Fig.~\ref{relative errors plot}. Although the scattering of the data is very large, the errors seem to roughly follow the predicted trends.

%\begin{figure*}[t]
%	\centering
%	\includegraphics[width=0.85\textwidth]{relative_errors.pdf}
%	\caption{Relative standard errors of $D_0$ (a) , $D_{\mathrm{PBC},zz}$ (b) and $\eta$ (b) 
%	 computed for varying system sizes. The data are shown
%	as a function of the number of molecules $N$ in the system.}
%	\label{relative errors plot}
%\end{figure*}

\subsection{Influence of the Simulation Length}
The length $\tau_\mathrm{block}$ of a simulation block, as reported in Table~\ref{systems}, was determined via the 
OrthoBoXY ``recipe'' introduced
in Ref.~\cite{buschComputingAccurateTrue2024}. 
This ``recipe'' allows the prediction
of consistent
accuracies of  $D_0$ and $\eta$ by taking into
account varying molecule sizes and the fluidity of the system.
The ``recipe'' is based on the simple fact that, according to the Einstein relation (Eq.~\ref{einstein method}), the three-dimensional mean squared displacement of a molecule during
a simulation run
%$\left< \Delta r^2 (\tau_\mathrm{block}) \right>$ 
%in the diffusive regime 
is described by
\begin{equation}
\left< |\Delta \mathbf{r}(\tau_\mathrm{block})|^2 \right> = 6 D \tau_\mathrm{block}\,.
\end{equation}
It was defined that each particle should on average have a displacement $(\langle|\Delta \mathbf{r}|^2\rangle)^{1/2}$ of 8.5 times its linear diameter 
$s\!=\!(V/N)^{1/3}$ per simulation block. The length $\tau_\mathrm{block}$ of the simulation block is given by
\begin{equation}
\label{orthoboxy recipe}
\tau_\mathrm{block} = \frac{(8.5\times s)^2}{6 D}\;.
\end{equation}
In hindsight, it might be, however, that the value
of $8.5\times s$ was very generous, such that
smaller simulation lengths might be perfectly adequate,
possibly leading to a significant reduction in required
computer resources.

\begin{figure}[t]
	\centering
	\includegraphics[width=0.4\textwidth]{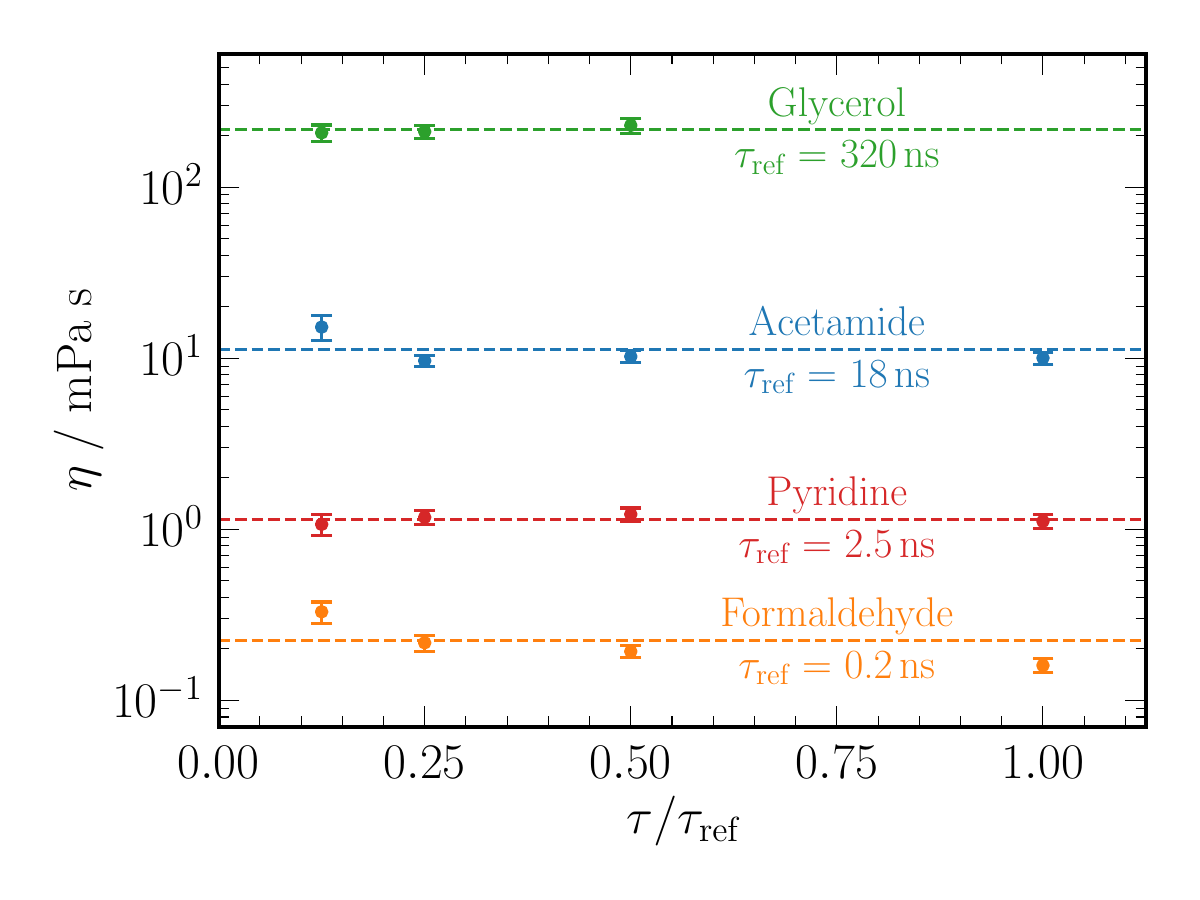}
	\caption{Viscosities of acetamide (blue), formaldehyde (orange), glycerol (green) and pyridine (red) with simulation block lengths scaled by factors of $1$, $\frac{1}{2}$, $\frac{1}{4}$ and $\frac{1}{8}$ (each with 40 simulation blocks). For glycerol, the simulation with a factor of $1$ was not performed. The dashed lines represent the averages.}
	\label{influence of t}
\end{figure}

For some of the investigated systems (Acetamide, Formaldehyde, Glycerol, and Pyridine), we have therefore performed simulations with block lengths $\tau_\mathrm{block}$ scaled by factors of $\frac{1}{2}$, $\frac{1}{4}$ and $\frac{1}{8}$ with respect to the reference length $\tau_\textrm{ref}$ defined by Eq.~\ref{orthoboxy recipe}. These scaled lengths correspond to a mean displacement of 6.0, 4.25, and 3.0 times the linear diameter, respectively. The viscosities from these simulations are displayed in Fig.~\ref{influence of t}. As can be seen there, the viscosities of glycerol (the system with the highest viscosity and highest $\tau_\textrm{ref}$) agree perfectly for the differently scaled values of $\tau_\mathrm{block}$. For acetamide and pyridine, the values also agree well, although the simulations with a scaling factor of $\frac{1}{8}$ seem to have higher uncertainties than the other simulations. For formaldehyde however, the viscosities are not constant for different values of $\tau_\mathrm{block}$. The calculated viscosity seems to increase as $\tau_\mathrm{block}$ becomes smaller, giving results that strongly deviate from the ``correct'' value. This may indicate that the particles do not show a fully diffusive behaviour in these shorter simulations.

As a consequence, we can safely recommend to reduce the
simulation block length for viscous systems to
$1/8\times\tau_\mathrm{block}$. For less
viscous systems a value of $1/4\times\tau_\mathrm{block}$
is suggested, while for highly fluid systems
$\tau_\mathrm{block}$ should not be shortened.

\section{Conclusions}

We calculated the shear viscosities of 15 neat molecular liquids from equilibrium MD simulations using the OrthoBoXY approach and compared them to viscosities calculated with the Green-Kubo method. The viscosities agree very well and demonstrate the validity of the OrthoBoXY approach. In addition, 
we point out numerical pitfalls for the computation of the viscosity
by discussing three variants for averaging the results. For
obtaining optimal results for the viscosity via Eq.~\ref{orthoboxy viscosity}
it is
recommended to utilize individual averages 
$\langle D_0\rangle\!=\!\langle D_{\mathrm{PBC},xx}+D_{\mathrm{PBC},yy}\rangle/2$
and $\langle D_{\mathrm{PBC},zz}\rangle$.
%We have derived an equation for the standard error of the viscosity using the propagation of uncertainty.

From simulations of multiple system sizes, we could verify that the viscosity of molecular liquids is not influenced by finite size effects for systems as small as 250 molecules. Additionally, we could show that also the standard error of the viscosity is nearly independent of the system size.
This is demonstrated to be a consequence
of a compensation effect of an increasing accuracy of the computed 
Einstein self-diffusion coefficients
with increasing systems-size and the system-size dependent weighting according to the OrthoBoXY-equation.

Finally, we revisited our previously introduced
``recipe'' for setting up
OrthoBoXY simulations with predictable numerical accuracy, suggesting  a certain simulation block length $\tau_\mathrm{block}$ based on the average displacement of the particles (see Eq.~\ref{orthoboxy recipe}). Based on data from simulations with different simulation lengths, we suggest the following: for highly viscous systems with large block lengths ($\tau_\mathrm{block} > 100\,\textrm{ns}$), the value of $\tau_\mathrm{block}$ might safely be scaled by a factor of $\frac{1}{8}$, significantly reducing the computational resources needed. For less viscous systemts with
medium block lengths ($1\,\textrm{ns} < \tau_\mathrm{block} < 100\,\textrm{ns}$), the value of $\tau_\mathrm{block}$ might safely be scaled by a factor of $\frac{1}{4}$ but caution is advised when going lower than that. For highly fluid systems
with short block lengths ($\tau_\mathrm{block} < 1\,\textrm{ns}$), the value of $\tau_\mathrm{block}$ should not be scaled in order to achieve reliable results. 
%Since these short simulations do not require many computational resources anyway, there is no need to sacrifice the accuracy of the data.
When combined with a smaller system size of 250 molecules,
these refinements are 
leading up to a 24-fold reduction in computational cost
compared to the previous recommended set-up
without sacrificing numerical
accuracy.

\section*{Supplementary Material}

The supplementary material provides self-diffusion data and computed 
viscosities for all studied systems as a function of subsequent MD simulation run blocks.

\begin{acknowledgments}
The authors thank the computer center at the University of
Rostock (ITMZ) for computational resources. 
R.L. is grateful to the Deutsche Forschungsgemeinschaft (DFG) for financing the projects LU 506/17-1, No. 470038970, and LU-506/18-1, po. 517661181.
\end{acknowledgments}

\section*{Author Declarations}

\subsection*{Conflicts of Interest}

The authors have no conflicts to disclose.

\subsection*{Author Contributions}

\textbf{Marcel Brandt}: 
Conceptualization (equal); 
Methodology (equal); 
Formal analysis (equal); 
Investigation (equal); 
Software (lead);
Data curation (lead);
Visualization (lead); 
Writing -- original draft (lead); 
Writing -- review \& editing (equal).
\textbf{Ralf Ludwig}: 
Conceptualization (supporting); 
Funding acquisition (lead); 
Supervision (supporting);
Writing -- review \& editing (supporting).
\textbf{Dietmar Paschek}: 
Conceptualization (equal); 
Methodology (equal); 
Formal analysis (equal); 
Investigation (equal); 
Software (supporting);
Data curation (supporting);
Visualization (supporting); 
Supervision (lead);
Project administration (lead);
Writing -- original draft (supporting); 
Writing -- review \& editing (equal).

\section*{Data Availability}

The code of GROMACS is freely available. The code of MDorado
is available via GitHub (\href{https://doi.org/10.5281/zenodo.21806953}{DOI: 10.5281/zenodo.21806953}).
Topology files, starting configurations and input parameters
for the MD simulations are available on Zenodo
(\href{https://doi.org/10.5281/zenodo.21805914}{DOI: 10.5281/zenodo.21805914}).

\section*{References}

%\printbibliography[title=References]

%\nocite{*}
\bibliography{lib}

%\bibliographystyle{rsc}

%\clearpage
\end{document}